\documentclass[aps,twocolumn,reprint,amsmath,amssymb,showpacs,superscriptaddress]{revtex4-1}
\usepackage{multirow}
\usepackage{graphicx,epsfig}
\usepackage{epstopdf}
\usepackage{wasysym}
\usepackage{stackengine}
\usepackage{color,soul} 
\usepackage{mathtools}
\usepackage{soul}
\usepackage[colorlinks=true,citecolor=blue,linkcolor=red,linktocpage=true,pagebackref=false]{hyperref}

\begin{document}
\newcommand{\be}{\begin{equation}}
\newcommand{\ee}{\end{equation}}
\newcommand{\ba}{\begin{eqnarray}}
\newcommand{\ea}{  \end{eqnarray}}
\newcommand{\ve}{\varepsilon}
\newcommand{\Tau}{\mathcal{T}}
\newcommand{\calb}{\mathcal{B}}
\newcommand{\calz}{\mathcal{Z}}
\newcommand{\tilH}{\tilde{H}}
\newcommand{\calp}{\mathcal{P}}
\newcommand{\cale}{\mathcal{E}}
\newcommand{\cald}{\mathcal{D}}
\title{Charge and energy transfer in ac-driven Coulomb-coupled double quantum dots}
\author{Mar\'{\i}a Florencia Ludovico}
\affiliation{Scuola Internazionale Superiore di Studi Avanzati (SISSA),
Via Bonomea 265, I-34136, Trieste, Italy}
\author{Massimo Capone}
\affiliation{Scuola Internazionale Superiore di Studi Avanzati (SISSA),
Via Bonomea 265, I-34136, Trieste, Italy}
\affiliation{Consiglio Nazionale delle Ricerche, Istituto Officina dei Materiali (IOM), Via Bonomea 265, I-34136, Trieste, Italy}

\begin{abstract}
We study the dynamics of charge and energy currents in a Coulomb-coupled double quantum dot system, when only one of the two dots is adiabatically driven by a time-periodic gate that modulates its energy level. 
Although the Coulomb coupling does not allow for electron transfer between the dots, it enables an exchange of energy between them which induces a time variation of charge in the undriven dot.
We describe the effect of electron interactions at low temperature using a time-dependent slave-spin 1 formulation within mean-field that efficiently captures the main effects of the strong correlations as well as the dynamical nature of the driving. We find that the currents induced in the undriven dot due to the mutual friction between inter-dot electrons are same order  than those generated in the adiabatically driven dot. Interestingly, up to 43$\%$ percent of the energy injected by the ac sources can be transferred from the driven dot to the undriven one. 
We complete our analysis by studying the impact of the Coulomb interaction on the resistance of the quantum dot that is driven by the gate.
\end{abstract}

\maketitle
\section{Introduction}
 
The study of transport through conductors coupled by the Coulomb interaction is a promising research field since the late 70's, when Pogrebinskii\cite{pogrebinskii} proposed an alternative way of measuring inner properties of solids which involved two electrically isolated 2D conductors (or layers) placed close together. The measurement protocol was based on the mutual friction (i.e. Coulomb-mediated scattering processes) felt by charge carriers belonging to different conductors due to long-range interactions. In these scattering processes momentum and energy can be transferred between the layers in spite of being electrically isolated from each other.

A spectacular effect of this mutual friction is the Coulomb drag\cite{pogrebinskii,drag2,reviewdrag}, in which a charge current is induced in an unbiased conductor (known as ``passive" conductor) simply applying a bias to the other (the ``active" conductor). This effect has been extensively studied in a wide variety of systems, from layered conductors\cite{bilayer1,bilayer2,bilayer3} to smaller dimensional systems as coupled quantum wires\cite{wiredot,wireexp1,wireexp2} or even double quantum dots structures\cite{qdots1,qdots3,qdots4}. In particular, experimental investigations in systems composed of Coulomb coupled
quantum dots are reported in Refs.\cite{qdots2,add1,add2,add3,add4,add5}.

Even more interestingly, not only charge currents but also heat and energy flows can be induced in the unbiased conductor thanks to the energy transfer between the Coulomb coupled objects. In recent years, this phenomenon has rekindled the interest of theoretical and experimental communities in Coulomb-coupled devices, especially in their thermodynamics, due to the possibility of using such energy transfer to develop novel nanotechnologies. Some examples are, the implementation of a single-electron heat diode\cite{diode}, a self contained quantum refrigerator \cite{qrefrigerador}, and the realization in a laboratory of a Szilard engine\cite{szilard}, an energy harvester \cite{add4,add5}, as well as an autonomous Maxwell's demon\cite{demon} that convert thermal energy into work by the use of information. Among the most recent works addressing the study of heat transport and entropy production in Coulomb coupled systems, we find Refs.\cite{heatqd1, heatqd3,heatqd2,heatqd4} for double quantum dot circuits, and Ref.\cite{tdrag} in the case of coupled Coulomb-blockaded metallic islands and quantum wires.

In this work we focus on several phenomena that boil down to a main question, namely what would happen if instead of a bias voltage or a thermal gradient, the active conductor is driven by time-dependent gates?  The natural questions regard the effects of the friction given by the Coulomb coupling on the scope of quantum pumping at low temperatures, the energy dissipation and the efficiency of the energy transfer between the Coulomb coupled dots.

\begin{figure}
 \includegraphics[width=0.41\textwidth]{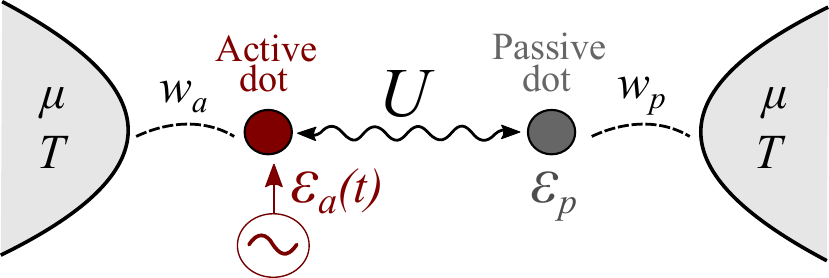}
  \caption{Scheme of the theoretical model considered in this work. It consists of two single level quantum dots, which are coupled in series with two non-interacting electron reservoirs with tunneling amplitudes $w_{a/p}$. The dots are coupled to each other through a Coulomb interaction of magnitude $U$, so that charge transfer between them is forbidden.
  The dot that is called ``active dot" is driven out of equilibrium by a time-periodic gate $\varepsilon_a(t)$, while the other quantum point (the ``passive dot") remains undriven with a constant energy level $\varepsilon_p$. All the reservoirs are at the same temperature $T$ and have the same chemical potential $\mu$.
  }\label{fig1}
\end{figure}

Here we make a first step towards the answer to these questions, which have not been discussed to the best of our knowledge. 
We believe that our results can shed light into the art of manipulating charge and energy fluxes, which is a crucial task for the development of new technologies. To achieve our goal we focused on the study of time-dependent charge and energy transport in a basic setup, which could be experimentally realized and is shown in Fig \ref{fig1}. It is composed by two Coulomb coupled quantum dots, namely the active dot and the passive dot, which are coupled in series with two electron reservoirs at the same temperature and chemical potential. Only the active quantum dot is driven by the application of an adiabatic time-periodic local gate that moves its level around the Fermi energy of the reservoir.

From the theoretical perspective, Coulomb-coupled quantum dot systems that are driven by bias voltages or a thermal gradients were previously studied mostly by the recourse of the master equation approach, valid in the regime where the hybridization with the reservoirs $\Gamma$ is negligible compared to the temperature and the Coulomb interaction $U$ \cite{qdots1,qdots3,qdots4,diode,qrefrigerador,heatqd1,tdrag}. Results were also presented by using the non-equilibrium non-crossing approximation (NCA) for rather high temperatures \cite{addnca}, since this approach fails at very low temperatures below the characteristic Kondo temperature. On the other hand, the use of master equations in similar systems and under the presence of adiabatic time-dependent drivings was addressed in Refs. \cite{addtdmeq1,addtdmeq2}.
At low temperatures and for a small interaction $U$, $U\ll\Gamma$, Ref.\cite{heatqd2} showed that the renormalized perturbation theory (RPT) in $U/\Gamma$ offers the most reliable description. However, in this work we focus on a different interesting regime, in which the temperature is very low and the interaction $U$ is larger than the hybridization.
In order to describe the adiabatically driven interacting system in this latter regime, we use the mean-field slave-spin 1 approach in Ref.\cite{slaveapin} that efficiently captures the main effects Coulomb coupling as well as the dynamical nature of the driving. As we discuss below, the slave-spin method realizes a mean-field which is suited to treat strong electron-electron interactions. 

%%%%%%%%%%
 %%%%%%%%%%

The paper is organized as follows. In Sec. \ref{model} we  introduce  the  model and the time-dependent slave-spin 1 approach within the adiabatic regime. Then in Sec. \ref{charge and energy} we compute charge and energy fluxes in the system.  Sec. \ref{results} presents the results for an illustrative example. Finally Section \ref{conclusions} is devoted to the summary and conclusions.

\section{Model and formalism}\label{model}

We consider the setup in Fig.\ref{fig1}, where we assume the quantum dots to be single-level with spinless electrons. This is actually a simplification, which has also been assumed before in the literature, see for instance Refs. \cite{qdots4,diode,heatqd2,heatqd4}, and that certainly could be experimentally realized by the application of a magnetic field able to completely polarize the dots. 

The prototypical device can be thought as composed by two subsystems, namely ``active" and ``passive", each containing a quantum dot and the corresponding reservoir. Accordingly, the active subsystem holds the active dot which is driven by the time-periodic local gate $\ve_a(t)$. The passive subsystem remains undriven, with the quantum dot at a constant energy level $\ve_p$. These two subsystems interact only through a Coulomb repulsion between inter-dot electrons, of magnitude $U$.      

Thus, we describe the full system by the Hamiltonian $H_{Full}(t)={\mathcal{H}}_a(t)+{\mathcal{H}}_p+Un_an_p$, where
\be
{\mathcal{H}}_\alpha={\mathcal{H}}^{dot}_\alpha+{\mathcal{H}}^{Res}_\alpha\;\;\;\text{for $\alpha=a,p$}
\ee
represents the uncoupled active and passive subsystems, with
\be\label{hamdot}
{\mathcal{H}}^{dot}_\alpha=\ve_\alpha n_\alpha+\sum_{\mathclap{k_\alpha}}w_\alpha(c^\dagger_{k_\alpha}d_\alpha +d_\alpha^\dagger c_{k_\alpha}),
\ee
for the quantum dots along with the tunneling couplings with the reservoirs, and
\be\label{ham}
{\mathcal{H}}^{Res}_\alpha=\sum_{\mathclap{k_\alpha}}\ve_{{k_\alpha}}c^\dagger_{k_\alpha}c_{k_\alpha},
\ee
for the non-interacting reservoirs, which are assumed to be at equilibrium.
The third term in $H_{Full}$ describes the Coulomb interaction, where the occupation operator reads ${n}_{a/p}=d^\dagger_{a/p}d_{a/p}$. On the other hand, the operator $c_{k_{a}}(c_{k_{p}})$ and its conjugate belong to the reservoir lying in the active (passive) subsystem, and $\ve_{k_a}(\ve_{k_b})$ is its energy band. The corresponding dot-reservoirs tunneling amplitudes
are $w_{a/b}$.

\begin{figure}
 \includegraphics[width=0.5\textwidth]{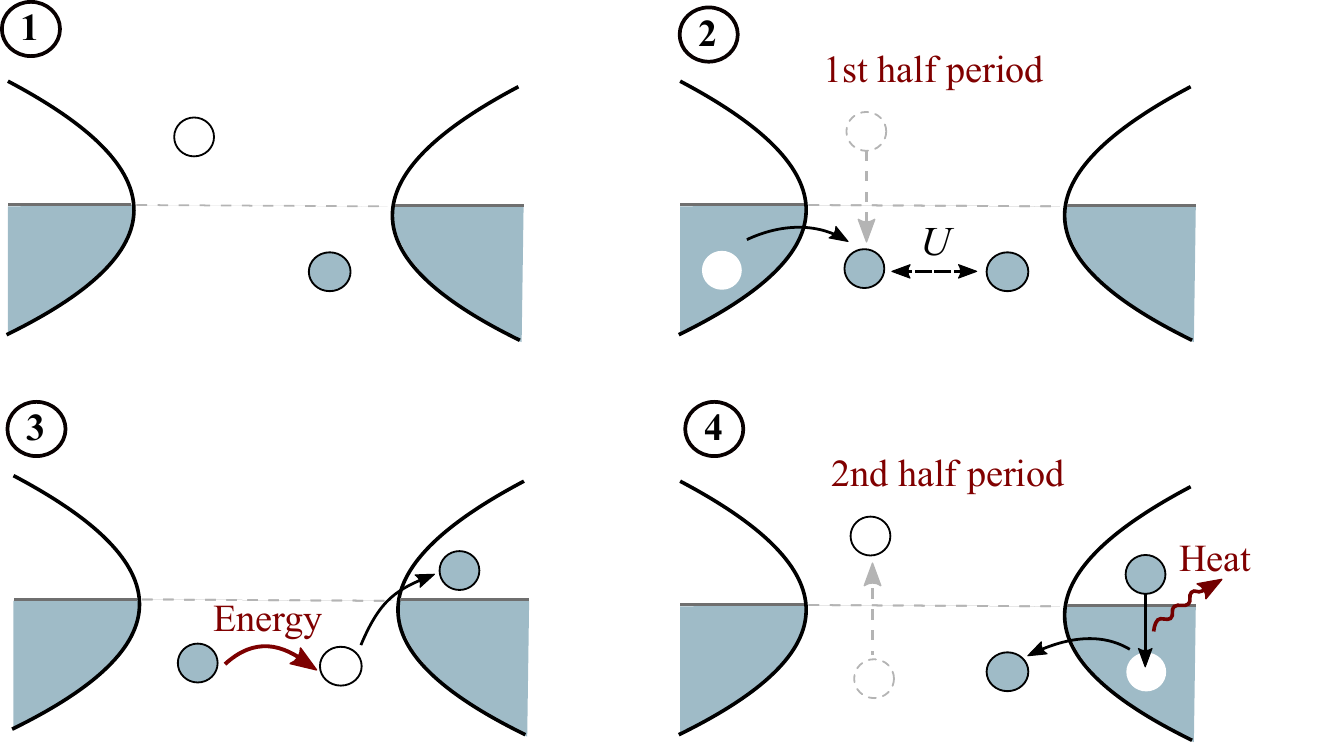}
  \caption{Scheme for energy transfer mechanisms. The active dot evolves as $\ve_a(t)=\ve_0\cos(\omega t)+\mu$ with $\ve_0>\mu$, while the levels of the passive dot levels remains constant and deep below $\mu$. 1)Initially, the active level is empty but the passive dot is filled. 2) During the 1st half of the oscillation period, the passive level goes below $\mu$, and in order to be filled, it has to pay the energy cost of the Coulomb repulsion. 3) Energy is transmitted from active to passive dots, so that the electron lying in the passive dot can tunnel above $\mu$. 4) During the 2nd half of the period, the active dot gets empty again so that the passive dot can return to be filled, dissipating heat during the process.
  }\label{fig2}
\end{figure}

The Coulomb coupling between the two quantum dots does not allow for an exchange of electrons between them. Therefore, energy transfer between active and passive subsystems will not be accompanied by any charge transfer between the two subsystems. Nevertheless, both quantum dots can certainly exchange particles with the reservoir they are attached to. In fact, as we are going to show later on in this section, energy transfer processes are accompanied (or mediated) by an induced time variation of charge in the passive dot. This charge current between the passive dot and the reservoir on the right is of zero net value when averaged over one oscillation period of the driving. 
Fig. \ref{fig2} shows in a more intuitive way how energy exchange processes take place. As an illustration, but without loss of generality, we consider an example in which all the reservoirs are at the same temperature $T=0$. The energy level of the active dot evolves as $\ve_a(t)=\ve_0\cos(\omega t)+\mu$ with $\ve_0>\mu$, while $\ve_p$ is constant and $\ve_p\ll\mu$. In this way, the initial configuration (see sketch 1) corresponds to the active level being above the Fermi sea while the passive dot lies below. Thus, the initial occupancy is $(n_a,n_p)=(0,1)$. During the first half period, in the second sketch, the active level fills up as it goes below $\mu$. Then, the Coulomb repulsion between electrons from different dots start to be felt, and this opposes to the filling of the active dot. Therefore, in order for the active dot to be filled up, it has to pay an energy cost which is extracted from the external time-dependent driving sources. Part of the energy that the active dot receives from the sources is then delivered to the passive dot so that the electron there can tunnel above the Fermi level of the right reservoir (see 3), leaving in this way the dot empty. By last, during the second half period of the driving, the emptying process of the active dot takes place and therefore the passive dot can be occupied again. Thus, as shown in step 4, an electron of energy $\ve_p$ from the right reservoir tunnels into the passive dot, generating a hole deep inside the Fermi sea. So then, an electron with an energy higher than $\mu$ decays to the hole, relaxing an amount of energy that will be then dissipated as heat in the reservoir but that could be eventually reused (or transformed) in a modified setup.

%%%%%%%%%%
 %%%%%%%%%%
 
Thus, we see that the driving of the active dot induces a time-dependent charge current between the passive dot and the reservoir on the right, which is neutral on average over a completed driving period. The passive dot has to receive some energy from the active subsystem in order to allow for these charge variations. This feature is properly captured by the time-dependent slave-spin 1 approach we previously introduced in Ref. \cite{slaveapin}, which is presented in the following section.

\subsection{Slave-spin 1 approach and the adiabatic regime}

We choose the slave-spin 1 (S-S1) mean-field approach for finite U introduced in Ref. \cite{slaveapin} as a minimal theoretical framework which captures the main effect of the electronic correlations in two-level systems, while also allowing for an analytical treatment that can be combined with a linear-response treatment for slow driving frequencies. Within the S-S1 approach, the original interacting Hamiltonian $H_{Full}(t)$ is represented in an enlarged Hilbert space that contains an auxiliary S=1 spin together with two pseudofermions.  The slave spin is in correspondence with the total fermionic number of the four possible electronic configurations for the double dot system $(n_a,n_p)=\{(0,0); (1,0); (0,1); (1,1)\}$.
Individual electrons are represented within this framework in terms of a pseudofermionic operator $d^*_{a/p}$ together with the $S=1$ spin. So that, operators belonging to the quantum dots are equally represented under the transformations: $d_{a/p}\rightarrow d^*_{a/p}S^-/(\hbar\sqrt{2})$ and $n_{a/p}\rightarrow n^*_{a/p}$, while the Coulomb interaction $n_an_p\rightarrow S_z(S_z+\hbar)/(2\hbar^2)$ can be rewritten in terms of the spin solely. The four physical electronic configurations are ensured by enforcing the following constraint on the total number of electrons 
\be\label{constraint}
n^*_a+n^*_p=\frac{S_z}{\hbar}+1.
\ee
Plugging the above transformations for the operators into Eq. (\ref{ham}), the S-S1 Hamiltonian of the full system can be written as
\ba\label{hamfull2}
H_{Full}^*(t)&=&\sum_{\alpha=a,p}{\mathcal{H}}^{{dot}^*}_{\alpha}(t)+{\mathcal{H}}^{Res}_\alpha\\
&&+\left(\frac{U}{2\hbar}S_z-\lambda(t)\right)\left(\frac{S_z}{\hbar}+1\right),\nonumber
\ea
with 
\ba\label{ham2}
\mathcal{H}^{{dot}^*}_\alpha(t)&=&\ve^*_\alpha(t){n}^*_\alpha+\sum_{\mathclap{k_{\alpha}}}\frac{w_{\alpha}}{\hbar\sqrt{2}}\left(S^-c^\dagger_{k_{\alpha}}d^*_\alpha+\text{H.c.}\right),
\ea
where $\lambda(t)$ is the Lagrange multiplier imposing the constraint on the occupancy in Eq. (\ref{constraint}) at every time, and $\ve^*_\alpha(t)=\ve_\alpha(t)+\lambda(t)$ are the re-normalized energy levels of the dots. In Eq. (\ref{hamfull2}), the Hamiltonians of the reservoirs $H^{Res}_\alpha$ remain the same since they are non-interacting, so that the S-S1 representation does not apply to them.   

Now, as customary in other slave-particles methods, we are going to treat the problem within the mean-field approximation (MF) that consist of decoupling fermionic and spin degrees of freedom and treating the constraint on average. The latter assumptions are justified for $U\gg \{\Gamma_{a/b}, \dot{\ve}_a\}$, where $\Gamma_{a/b}$ are the hybridizations with the active and passive reservoirs. Thereby, fluctuations of the spin with respect to the mean values can be neglected even under the action of the time-dependent driving, as long as the adiabatic condition $\dot{\ve}_a\ll \Gamma_{a/b}$ is satisfied. Then, we should replace the components of the salve-spin $S_z$, $S^{+/-}=S_x \pm i S_y$ with their expectation values $\langle S \rangle$, and neglect their fluctuations. Consequently, the interacting Hamiltonian $H_{Full}(t)$ is mapped into a non-interacting one
\be\label{hammf}
\tilde{H}_{full}(t)=\tilde{\mathcal{H}}_a(t)+\tilde{\mathcal{H}}_p(t)+\beta(t),
\ee
where $\tilde{\mathcal{H}}_\alpha(t)$ is the effective MF Hamiltonian for the subsystem $\alpha=\{a,p\}$, with $\tilde{\mathcal{H}}_\alpha(t)=\tilde{\mathcal{H}}_\alpha^{dot}(t)+{\mathcal{H}}_{\alpha}^{Res}$,
\be\label{hamdotmf}
\tilde{\mathcal{H}}_{\alpha}^{dot}(t)=\ve^*_\alpha(t){n}^*_\alpha+\sum_{\mathclap{k_{\alpha}}}{w^*_{\alpha}(t)}c^\dagger_{k_{\alpha}}d^*_\alpha+\mbox{H.c.},
\ee
and
\be\label{betaoriginal}
\beta=\frac{U}{2\hbar^2}\left(\langle S^2_z\rangle+\hbar\langle S_z\rangle\right)-\frac{\lambda}{\hbar}\left(\langle S_z\rangle+\hbar\right),
\ee 
with the re-normalized tunneling factors being $w_\alpha^*(t)=w_\alpha\langle S^-\rangle(t)/(\hbar\sqrt{2})$. Then, we can see that within the mean-field S-S1 framework ``active" and ``passive" subsystems are described as effectively uncoupled from each other, even though they actually interact via the Coulomb repulsion. In fact, all the information about the interaction between the subsystems is contained in 
the effective Hamiltonian parameters $\ve_{a/p}^*(t)$ and $w_{a/p}^*(t)$, which are all time-periodic functions due to the periodicity of the driving. Although in the real setup the driving is merely applied to the active dot, now under the mean-field S-S1 both the dots turn out to be driven in time. This is because the Coulomb interaction is portrayed as extra time-dependent driving sources, acting locally on the levels of the dot as well as on the contact with the leads. The latter feature of this model is advantageous for describing the energy transfer mechanisms sketched in Fig. \ref{fig2}, since the induction of transport in the passive subsystem due to the interaction is simply explained in terms of extra time-dependent driving sources acting on that part of the device. 

For finding the effective Hamiltonian parameters $\ve^*_{a/p}(t)$, $w^*_{a/p}(t)$, and the function $\beta(t)$, the coupled problem between fermionic and spin dynamics must be solved.
In this work we consider the driving to be within the adiabatic regime ($ad$), namely a slow evolution in time of the parameter $\dot{\ve}_a\rightarrow 0$. For this regime,  as explained in detail in Ref. \cite{slaveapin}, the spin dynamics is simplified since all the spin components turn out to depend on $\langle S_z \rangle$ solely, as the only independent component. In particular, for the Hamiltonian parameters we have
\ba\label{relationsS}
&&\vert\langle S^-\rangle^{ad}\vert^2=\hbar^2-\langle {S_z\rangle^{ad}}^2,\\
&&\langle S_z^2\rangle^{ad}=({\langle S_z\rangle^{ad}}^2+\hbar^2)/2.\nonumber
\ea 
In this way, the vertical component $\langle S_z\rangle^{ad}(t)$ together with $\lambda^{ad}(t)$ constitute the full set of variables describing the the Coulomb interaction, and their dynamics is obtained by solving the following set of slave-spin equations \cite{slaveapin} composed by:   
The equation of motion for $S_z$, that is
\ba\label{consteqad}
0&=&\left(\lambda^{ad}-\frac{U}{2}\langle n^*\rangle\right)\left(1-\frac{{ \langle S_z\rangle^{ad}}^2}{\hbar^2}\right)
\nonumber\\
&&+\frac{ \langle S_z\rangle^{ad}}{\hbar}\sum_{\mathclap{\alpha=a,p}}2\mbox{Re}\left\{w^*_\alpha\langle c^\dagger_{k_\alpha}d^*_\alpha\rangle\right\},
\ea
with $n^*=n^*_a+n^*_p$,
and the constraint on average
\be\label{eq12eqad}
\frac{\langle S_z\rangle^{ad}}{\hbar}+1=\langle n^*\rangle=\sum_{\alpha=a,p}\langle {d^*_\alpha}^\dagger d^*_\alpha\rangle.
\ee
In the above equations the expectation values for the pseudo-fermions $\langle n^*\rangle$ and $\langle c^\dagger_{k_\alpha} d^*_\alpha\rangle$ should be consistently evaluated within the adiabatic regime, namely in linear response in the small variation of the effective parameters $\dot{\ve}^*_\alpha$ and $\dot{w}_\alpha^*$, which involve the time-derivatives of the variables $\dot{\lambda}^{ad}$ and ${\dot{\langle S_z\rangle}}^{ad}$ (for further details see Ref. \cite{slaveapin}). Then it is important to notice that even though Eqs. (\ref{consteqad}) and (\ref{eq12eqad}) appear to be stationary, they actually  constitute a system of ordinary differential equations due to the presence of time-derivatives of the variables inside the electronic expectation values. Although the above set of equations could be difficult to solve, the quasi-static evolution of the system makes it possible to approximate the solutions as little variations around the static (or frozen) solutions at every time $t$:
\ba\label{solap}
\langle S_z\rangle^{ad}(t)&\sim & S_z^{t}+\delta S_z\\
\lambda^{ad}(t)&\sim &\lambda^{t}+\delta \lambda,\nonumber
\ea
where the index $t$ means that they are static values, in the sense that the dependence on time is purely parametric, as in a series of snapshots of the system in equilibrium with frozen parameters. The first order corrections taking into account the effect of the slow driving are $\delta S_z$
and $\delta\lambda$, which depend on the frozen solutions and are $\propto \dot{\ve}_a$. The expansion in Eq. (\ref{solap}) has the advantage of offering a practical description of the dynamics in terms of the frozen static values $S_z^t$ and $\lambda^t$ solely, for which Eqs. (\ref{consteqad}) and (\ref{eq12eqad})
are reduced to a stationary system of non-linear equations as presented in Ref.\cite{slaveapin}. When evaluated in the system model of this work, the latter set stationary equations reads
\ba\label{consteq}
0&=&\left[\lambda^t-\frac{U}{2}\left(1+\frac{ S_z^t}{\hbar}\right)\right]\left(1-\frac{{ S_z^t}^2}{\hbar^2}\right)\\
&&+\frac{ S_z^t}{\hbar}\sum_{\mathclap{\alpha=a,p}}\;\int\frac{d\varepsilon}{\pi}{\rho}^t_\alpha(\varepsilon)(\varepsilon-{\ve^t_\alpha})f(\varepsilon),\nonumber
\ea
and
\be\label{eq12eq}
\frac{S_z^t}{\hbar}+1=\sum_{\alpha=a,p}\int\frac{d\varepsilon}{2\pi}{\rho}_\alpha^t(\varepsilon)f(\varepsilon).
\ee
Here ${\ve^t_\alpha}=\ve_\alpha(t)+\lambda^t$, and $\rho_\alpha^t(\ve)=\Gamma_\alpha^t/[(\ve-\ve^t_\alpha(t))+i\Gamma_\alpha^t/2]$, with $\Gamma_\alpha^t=\Gamma_\alpha(\hbar^2-{S_z^t}^2)/2\hbar^2$ being the effective hybridization, is the density of states of the quantum dot in subsystem $\alpha$. We are considering the wide-band limit, where the bare hybridization reads $\Gamma_\alpha=w_\alpha^2{\theta}$ with $\theta$ being the energy independent density of states of the $\alpha$-lead. On the other hand, $f(\ve)$ is the Fermi-Dirac distribution which in this work is the same for both reservoirs. 
Linear response terms in Eqs. (\ref{consteqad}) and (\ref{eq12eqad}) are taken into account when computing the corrections $\delta S_z$ and $\delta\lambda$, which as mentioned above, depend on the above frozen solutions and also on the spectral properties of the system. These latter can be found by solving a simple system of linear equations, but since they are not crucial for this work, we refer the reader to Ref. \cite{slaveapin} for details on how to compute them.

\section{Charge and energy fluxes}\label{charge and energy}
\subsection{Pumping charge}
In this work we focus on the effect of the Coulomb coupling on charge and energy transfer, with a particular interest in the passive subsystem, where the transport is exclusively induced by the action of $U$. 

Regarding the transport of charge, electronic pumping currents flow in the contacts with the reservoirs in response to the time-periodic driving, while there cannot be any exchange of particles between the two dots. Then, the charge current entering reservoir $\alpha$, $I_\alpha(t)$, should obey the following charge conservation law
\be
I_\alpha=-e\frac{d\langle n^*_\alpha\rangle}{dt},
\ee
that establishes a relation between the currents entering the reservoirs and the charge variations in the dots.  
The lowest-order contribution of the above current, $I^{(1)}_\alpha \equiv I_\alpha^{pump}$, is of  first order in the variation of the Hamiltonian parameters, $\dot{\ve^*_\alpha}(t)$ and $\dot{w_\alpha^*}(t)$ that are $\propto\hbar\omega$. In Ref.\cite{motor}, we have already provided an expression for the first-order (or linear response) current in the case of a single non-interacting quantum dot being driven by time-dependent couplings to the leads and also a time-dependent energy level. These calculations are applicable to the present problem within the mean-field approximation, for which active and passive dots appear to be non-interacting and uncoupled to each other. Then the current must be evaluated with the effective parameters given by the S-S1 mean-field approach, which contain all the information about the finite $U$ coupling. Consequently, the pumping current reads 
\ba\label{curr}
\frac{I_\alpha^{pump}}{e}&&=\sum_{k_\alpha}\frac{i}{\hbar}\langle\left[\tilde{H}_{full},c_{k_\alpha}^\dagger c_{k_\alpha}\right]\rangle\nonumber\\
&&=\int\frac{d\ve}{2\pi}\frac{df}{d\ve}\rho_\alpha^t\Gamma_\alpha^t\partial_t\left(\frac{\ve-\ve_\alpha^t}{\Gamma_\alpha^t}\right).
\ea

\subsection{Power and energy transfer}

The Coulomb interaction does not allow for an exchange of electrons between the two quantum dots. Yet, it allows for a net energy transfer, as sketched before in Fig. \ref{fig2}.
In order to study the latter energy transfer mechanisms between active and passive subsystems, we should first analyze the variation of the energy in the full system and then the way it is distributed in the two subsystems. In  contrast to the pumped charge, that is conserved for the full system, the corresponding rate of change in the total energy is equal to the power developed by the external ac-sources
\be\label{power1}
P^{ac}(t)=\langle \partial_t H_{full}\rangle=\dot{\ve}_a(t)\langle n_a\rangle=\dot{\ve}_a(t)\langle n_a^*\rangle.
\ee
At this point, although $\langle n^*_a\rangle(t)$ could be simply computed as the time integral of the current in Eq. (\ref{curr}) for $\alpha=a$, it turns out to be more convenient to work with the effective Hamiltonian, as $P^{ac}=\langle \partial_t \tilde{H}_{full}\rangle$, because this allows us to easily identify how the energy is distributed between the two subsystems during the driving. However, at first glance, the fact that the mean-field S-S1 introduces extra time-periodic parameters for describing the interaction, implies that some care should be taken before replacing  the original Hamiltonian $H_{full}$ by the approximated $\tilde{H}_{full}$ in the definition of the power as it is usually done for computing the charge current{\footnote{We stress here that transforming the Hamiltonian under the S-S1 framework $H_{full}\rightarrow H_{full}^*$ is an exact representation, while approximations are imposed only when treating the problem within mean-field with the Hamiltonian $\tilde{H}_{full}$.}}. However, in our case the substitution is justified since (as we will show in the following)
\be\label{eqH}
\langle \partial_t H_{full}\rangle^{ad}=\langle \partial_t \tilde{H}_{full}\rangle^{ad}
\ee
at least within the adiabatic regime. 

In order to prove the latter equality, we start by comparing Eq. (\ref{power1}) with the time derivative of the MF Hamiltonian in Eqs. (\ref{hammf}) and (\ref{hamdotmf})  
\ba
\dot{\ve}_a(t)\langle n^*_a\rangle&=&\sum_{\alpha=a,p}\dot{\ve}_\alpha^*(t)\langle n^*_\alpha\rangle +2\mbox{Re}\left\{\dot{w}_\alpha^*(t) \langle c^\dagger_{k_\alpha} d^*_\alpha\rangle\right\}\nonumber\\
&&+\dot{\beta}(t),
\ea
and since $\dot{\ve}^*_a(t)=\dot{\ve}_a(t)+\dot{\lambda}(t)$, $\dot{\ve}^*_p(t)=\dot{\lambda}(t)$  and $\dot{w}_\alpha^*(t)={w_\alpha}d_t\langle S^-\rangle/{\hbar\sqrt{2}}$, the previous equation is then reduced to 
\be\label{betadot}
\dot{\beta}(t)=-\dot{\lambda}(t)\langle n^*\rangle\!-\!\!\!\sum_{\alpha=a,p}\!\!\frac{\sqrt{2} w_\alpha }{\hbar}\mbox{Re}\left\{\dot{\langle S^-\rangle}\langle c^\dagger_{k_\alpha} d^*_\alpha\rangle\right\}.
\ee
Now, taking Eq. (\ref{betaoriginal}) and using the relations between the components of the spin in (\ref{relationsS}), we can express $\beta$ in the adiabatic regime as
\be\label{bad}
\dot{\beta}^{ad}=\left(\frac{U}{2}\langle n^*\rangle-\lambda^{ad}\right)\dot{\langle S_z\rangle}^{ad}-\dot{\lambda}^{ad}\langle n^*\rangle,
\ee
and, on the other hand, we also know from the time derivative of Eq. (\ref{relationsS}) that
\be\label{ders}
{\dot{\langle S^-\rangle}}^{ad}=-\frac{\hbar\langle S_z\rangle^{ad}}{\langle S^-\rangle^{ad}}{\dot{\langle S_z\rangle}^{ad}}.
\ee
Now, plugging Eqs. (\ref{bad}) and (\ref{ders}) into (\ref{betadot}), we obtain 
\be
\left(\frac{U}{2}\langle n^*\rangle^{ad}-\lambda^{ad}\right)=\frac{\hbar\langle S_z\rangle^{ad}}{\vert \langle S^-\rangle^{ad}\vert^2}\!\sum_{\alpha=a,p}\!2\mbox{Re}\left\{\! w_\alpha^*\langle c^\dagger_{k_\alpha} d_\alpha^*\rangle\right\},
\ee
that it is the same as the slave-spin equation (\ref{consteqad}), which shows that Eq. (\ref{eqH}) is satisfied. 

\subsubsection{Energy distribution}

Now that we have already shown the validity of Eq. (\ref{eqH}), so that the MF preserves the definition of the power $P^{ac}=\langle\partial_t\tilde{H}_{full}\rangle^{ad}$, we can start analyzing the energy distribution in the system. From Eq. (\ref{hammf}) we know that
\be\label{power}
P^{ac}=\frac{d}{dt}\langle\tilde{\mathcal{H}}_a\rangle +\frac{d}{dt}\langle\tilde{\mathcal{H}}_p\rangle+\dot{\beta},
\ee
where the first two terms tell us that a portion of the energy delivered by the external ac-sources is temporally stored in the active and passive subsystems, while there is also a third contribution from the temporal variation of function $\beta$. This latter function, that is exclusively introduced by the MF, it is constant for stationary systems and so it is
generally discarded. Nonetheless, it may not be dismissed when studying time-resolved energy transfer because its dynamical nature impacts on the power. Due to the periodicity of the MF parameters $\lambda$ and $\langle S_z\rangle$ in Eq.(\ref{betaoriginal}), $\beta$ turns out to be also a time-periodic function and, as such, its time-derivative vanishes when averaged over one driving period $\tau=2\pi/\omega$, $\overline{\dot{\beta}}\tau=\int_0^\tau \dot{\beta}dt=\beta(\tau)-\beta(0)=0$. 
Thus, we identify $\dot{\beta}$ as a conservative term in the power insomuch as it does not give a net contribution to the rate of change of the energy.

On the other hand, the energy rates $d {\langle {\tilH}_{\alpha}\rangle}/dt$ with $\alpha=\{a,p\}$ are exactly the power $P_\alpha$ developed by the effective potentials in the subsystem $\alpha$,
\be\label{powersub}
P_\alpha=\frac{d}{dt}\langle\tilde{\mathcal{H}}_\alpha\rangle=\dot{\ve}_\alpha^*(t)\langle n^*_\alpha\rangle +2\mbox{Re}\left\{\dot{w}_\alpha^*(t) \langle c^\dagger_{k_\alpha} d^*_\alpha\rangle\right\}.
\ee
That will have a conservative contribution, $P^{cons}_\alpha$, and one dissipative  $P_\alpha^{diss}$, so that $P_\alpha=P^{cons}_\alpha+P^{diss}_\alpha$. As explained in Ref. {\cite{singledot}} for a single dot device, the conservative component $P^{cons}_\alpha$ corresponds to an amount of energy that is temporally stored in the quantum dot with a zero net value $\overline{P^{cons}_\alpha}=0$, and it is not related to the energy transferred to the reservoirs, which is purely dissipative and therefore it is contained in $P_\alpha^{diss}$. Particularly in this work, we will focus on the dissipate components of the power in Eq. (\ref{power}), that are those giving a net transfer of energy from the active to the passive subsystem. 
For evaluating the corresponding dissipative component of the power in Eq. (\ref{powersub}), we again follow the procedure of Ref. \cite{motor}, to which we refer the reader for further details. Thus,
\be\label{powerdiss}
P^{diss}_\alpha=\hbar\int\frac{d\ve}{4\pi}\frac{df}{d\ve}\left(\rho_\alpha^t\Gamma_\alpha^t\partial_t\left(\frac{\ve-\ve_\alpha^t}{\Gamma_\alpha^t}\right)\right)^2.
\ee
The dissipation of energy in each subsystem will be in the form of heat deep inside the reservoirs $\alpha$, so that  $P^{diss}_\alpha=\dot{Q}_\alpha$. However, due to the arrangement of the quantum dots in the device, the passive subsystem can get energy only from the active subsystem. In the way that all the heat that is dissipated in the passive reservoir must correspond to energy coming from the active subsystem. Consequently, we may identify the heat $\dot{Q}_p$ in passive reservoir as the flux of energy exchanged between the subsystems, $\dot{E}^{a\rightarrow p}\equiv\dot{Q}_p$ that goes from the active to the passive subsystem.  

In this way, the equation for the net energy distribution in the full system reads  
\be\label{energydistribution}
\overline{P^{ac}}=\overline{P^{ac}_{diss}}=\overline{P_a^{diss}}+\overline{P_p^{diss}}=\overline{\dot{Q}_a}+\overline{\dot{E}^{a\rightarrow p}},
\ee
and shows that the energy flux $\overline{{P}^{ac}}$ injected in the active subsystem is partly dissipated as heat $\overline{\dot{Q}_a}$ in reservoir $a$, while the rest $\overline{\dot{E}^{a\rightarrow p}}$ is transferred to the passive subsystem. In our setup, this latter energy that is exchanged between the subsystems is then dissipated as heat, but
could be eventually transformed into useful work in a proper device.

\section{Results}\label{results}

As an illustrative example for the energy transfer mechanisms in Fig. \ref{fig2}, we consider the case $\ve_a(t)=\ve_0\cos(\omega t)+\mu$ within the adiabatic regime, so that $\hbar\omega\ll\Gamma_\alpha$, while $\ve_p$ being a constant function. In particular, we focus on the situation in which the system is at temperature $T= 0$, and with balanced hybridizations with the reservoirs $\Gamma_a=\Gamma_p=\Gamma/2$. In what follows, we study the dynamics of charge and energy fluxes for different intensities of the Coulomb interaction $U$ and various values of the energy level of the passive dot $\ve_p$.

As was shown in the previous section, the pumping charge currents $I_\alpha^{pump}$ in Eq. (\ref{curr}), as well as the dissipative components of the power $P^{diss}_\alpha$ in Eq. (\ref{powerdiss}) are evaluated only at the frozen parameters $\lambda^t$ and $S_z^t$. The latter can be found by solving the non-linear system of stationary equations in Eqs. (\ref{consteq}) and (\ref{eq12eq}) at every instant of time $t$.  We emphasize that the frozen picture used here considers the system to be at equilibrium at every time $t$ as in a sequence of snapshots, in the way that the time variable is treated as a parameter. 

\begin{figure}
 \includegraphics[width=0.5\textwidth]{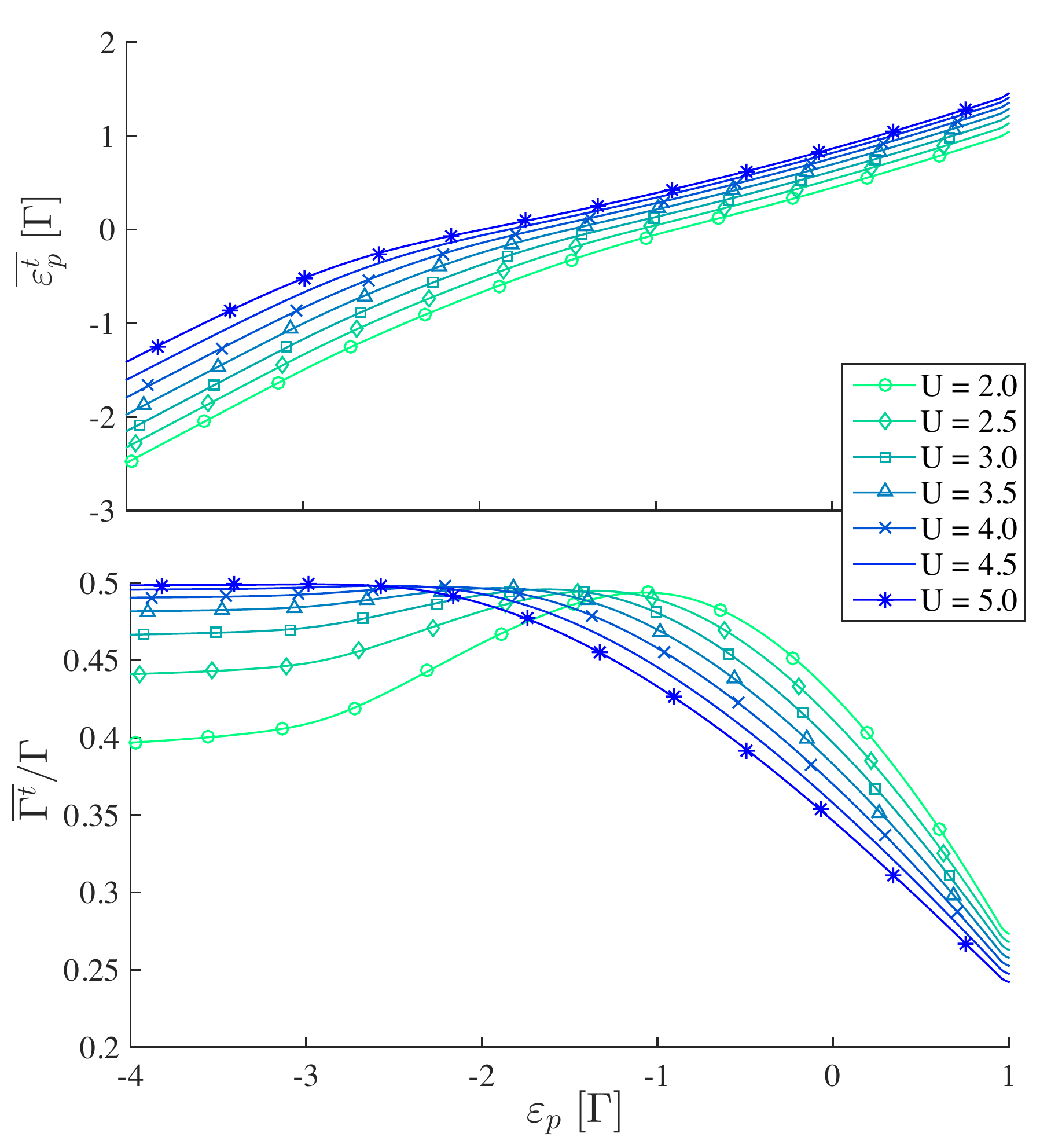}
  \caption{Averaged values over a driving period $\tau$ of the effective energy level of the passive dot $\overline{\ve_p^t}$ (top panel) and the total effective hybridization $\overline{\Gamma^t}$ (bottom panel) as functions of the energy level of the passive dot $\ve_p$, for different values of the interaction $U$.   
  The parameters are: $\ve_0=3\Gamma$, $\mu=0$, $\hbar\omega=10^{-3}\Gamma$, and $T=0$. All the energies, including the values of $U$ inside the legend, are expressed in units of $\Gamma$.
  }\label{fig6}
\end{figure}

In Fig. \ref{fig6} we show the average values over a single driving period of the frozen effective level of the passive dot $\overline{\ve_p^t}=\ve_p+\overline{\lambda^t}$, together with the re-normalizing factor of the hybridizations $\overline{\Gamma^t}/\Gamma=\overline{\Gamma^t_\alpha}/\Gamma_\alpha=(\hbar^2-\overline{{S_z^t}^2})/2\hbar^2$, which is the same for each of the reservoirs $\alpha=\{a,p\}$ as for the total $\Gamma^t=\Gamma^t_a+\Gamma^t_b$. Here, we vary $\ve_p$ within a range of energies in which the passive quantum dot is always occupied. From $\ve_p\ll\mu$ where the average occupation $\overline{\langle n_p^*\rangle}\rightarrow 1$, to the limit $\ve_p\rightarrow \Gamma$ in which the passive dot is almost empty $\overline{\langle n_p^*\rangle} \sim 0.05$ but still occupied. Beyond this range of energies, for $\ve_p>\Gamma$, the un-driven dot is empty thus the Coulomb interaction vanishes, and so does the effect we want to study.    

%%%%%%%%%%
 %%%%%%%%%%
 
 As it is already known, the interaction $U$ has the effect of moving up the resonance from its non-interacting value $\mu$ (we set $\mu=0$), and this upward shift is represented in our model by the Lagrange multiplier $\lambda^t$ that is always a positive number \cite{slaveapin}. This effect can be seen in the top panel of Fig. \ref{fig6} from the fact that the effective level of the passive dot $\overline{\ve_p^t}$ presents a change of sign (i.e. it crosses the Fermi level $\mu=0$) at a critical energy $\ve_p^c(U)$, such that $\overline{\ve_p^t}(\ve_p^c)=0$, which is smaller than the chemical potential $\ve_p^c(U)<\mu$. The critical energy depends on the strength of the Coulomb interaction $U$, so that the crossing occurs earlier for larger values of $U$ than for lower values of the interaction. As expected, we notice that the shift in the energy levels $\overline{\lambda^t}$ increases as the strength of the Coulomb interaction rises up, so the curves for higher values of $U$ are above those corresponding to a lower $U$. Moreover, a slope change in the curves of $\overline{\ve^t_p}$ can be perceived  around the critical values $\ve_p^c(U)$.
 When the effective level of the passive dot is deep below the Fermi energy $\ve_p\ll\ve_p^c$, the slope $\gamma=1+d\overline{\lambda^t}/d\ve_p\sim 1$, which means that the Lagrange multiplier is approximately a constant function of $\ve_p$. On the contrary, $0<\gamma<1$ and then $d\overline{\lambda^t}/d\ve_p<0$ as the effective level moves further above the resonance $\ve_p\gg\ve_p^c$, which tells us that the average energy shift $\overline{\lambda^t}$ decreases as the occupation of the passive dot gets diminished.

Now we turn to the behaviour of the mean value of the total frozen hybridization $\overline{\Gamma^t}$, which provides information about the electronic configuration of the double-dot system. 
As explained in the previous sections, the component of the spin $S_z$ is in correspondence with the total occupation number $n_a^*+n_p^*$ through
the constraint in Eq. (\ref{constraint}).  In this way, we know that the two-level system is empty when $S^t_z=-\hbar$, it is double occupied for $S^t_z=\hbar$, and it is filled with a single electron when $S_z^t=0$. Therefore, the effective hybridization is $\Gamma^t=0$ when the system is either double occupied or empty, while it attains its maximum value $\Gamma^t=\Gamma/2$ at the single occupied state. Results are shown in the bottom panel of Fig. \ref{fig6}. We can see that when the  level of the passive dot is deep below the Fermi sea $\ve_p\ll\mu$, the double dot-system approaches the single occupied state $\overline{\Gamma^t}\rightarrow \Gamma/2$ as the intensity of $U$ increases. This is simply because in this limit $\overline{\langle n_p^*\rangle}\rightarrow 1$ and the repulsive Coulomb interaction prevents the system from the double occupancy, so that for high values of $U$ the active dot shall be almost empty on average $\overline{\langle n_a^*\rangle}\sim 0$. Whereas, when the intensity of $U$ decreases, both the dots could be simultaneously occupied for some time-interval in the oscillation period. And this leads $\overline{\Gamma^t}$ to take smaller values. 

On the other hand, we observe a significant change in the behaviour of $\overline{\Gamma^t}$ as the effective passive level overcomes the Fermi energy, for $\ve_p>\ve_p^c(U)$. Now, smaller values of $U$ lead to higher values of the hybridization. Moreover, $\overline{\Gamma^t}$ starts to decrease until it reaches half its maximum value $\overline{\Gamma^t}\sim 0.25\Gamma$ for $U=5\Gamma$ when $\ve_p\rightarrow \Gamma$, which means that the system is single occupied for half the oscillation period while it is empty for the rest of the time. The latter features can be understood as follows. As the effective level of the passive dot moves further above $\mu$, its average occupancy decreases, and so does the average $\overline{\lambda^t}$. In the limit $\ve_p\rightarrow \Gamma$ and for large values of $U$, the passive dot is almost empty, while the level of the active dot can oscillate between being occupied and empty $\overline{\langle n_a^*\rangle}\sim 0.45$. As the strength of the interaction $U$ is reduced, the upward shift in the energies $\overline{\lambda^t}$ gets smaller thus the effective levels are closer to $\mu$ and then they are more occupied on average. This increases the total occupation of the double-system and therefore $\overline{\Gamma^t}$ rises up, which explains the fact that the curves for higher values of $U$ are below those corresponding to a less intense interaction.

\subsection{Charge current}
%%%%%%%%%%
\begin{figure}
 \includegraphics[width=0.5\textwidth]{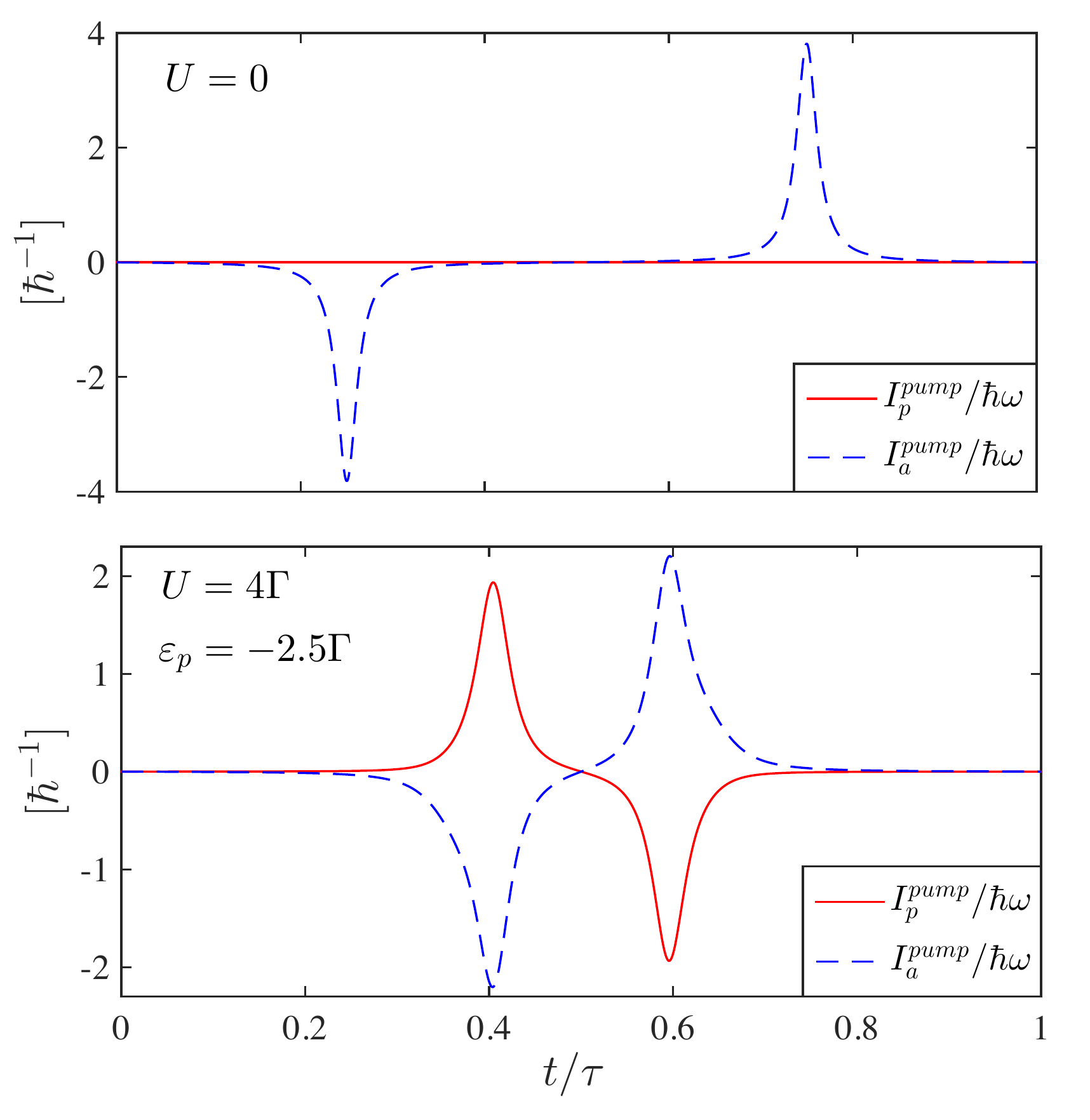}
  \caption{Linear response coefficients of the charge currents $I_a^{pump}/\hbar\omega$ and $I_a^{pump}/\hbar\omega$, which are $\omega$-independent, as a function of time. Top panel: Non-interacting limit, $U=0$. Bottom panel: results for $U=4\Gamma$ and $\ve_p=-2.5\Gamma$. Other parameters are the same as in Fig. \ref{fig6}. 
  }\label{fig3}
\end{figure}
 %%%%%%%%%%

We now turn to analyze the linear response charge current in Eq. (\ref{curr}) that is pumped into the reservoirs. As an example, Fig. \ref{fig3} shows $I^{pump}_a$ and $I^{pump}_p$ as functions of time for $U=4\Gamma$ and $\ve_p<\mu$, as well as when $U=0$ for which the quantum dots are disconnected from each other. We show that, as explained before, currents in the passive dot are merely induced by the finite interaction $U$, thus $I^{pump}_p\vert_{U=0}=0$. Naturally, the peaks in the charge currents $I^{pump}_\alpha$ occur within a time interval in which $\vert\ve^t_\alpha-\mu\vert\lesssim\Gamma_{\alpha}^t$, but interestingly we note that charge fluctuations in the two subsystems are completely synchronized (i.e. opposite in sign). So that, during the first half period, an electron leaves the passive dot and enters the right reservoir (positive peak) at the same time when an electron from the reservoir on the left is entering the active dot (negative peak). Then the process is reversed during the second half period. Moreover, we find that the induced currents in the undriven passive dot are just smaller but same order (first order in the variation of the driving $\dot{\ve_a}$) than those generated in the active dot. This is different from what was observed in Coulomb drag devices, for which drag currents are in general second order in the applied perturbation, i.e. bias voltage or thermal gradient \cite{qdots2,qdots4,diode,heatqd2,tdrag}. Something surprising is that currents are induced in the passive dot even at zero temperature, which evidence that unlike drag currents that are $\sim T^2$, the induction of transport in the passive part  can not be interpreted as a rectification of thermal fluctuations \cite{cdrag}.   
On the other hand, we also observe a reduction in amplitude of $I^{pump}_a$ at finite $U$ with respect to the non-interacting case, which is due to the mutual friction between inter-dot electrons (i.e. Coulomb mediated scattering processes) that opposes the filling of the active dot when the passive dot is occupied.   

\begin{figure}
 \includegraphics[width=0.5\textwidth]{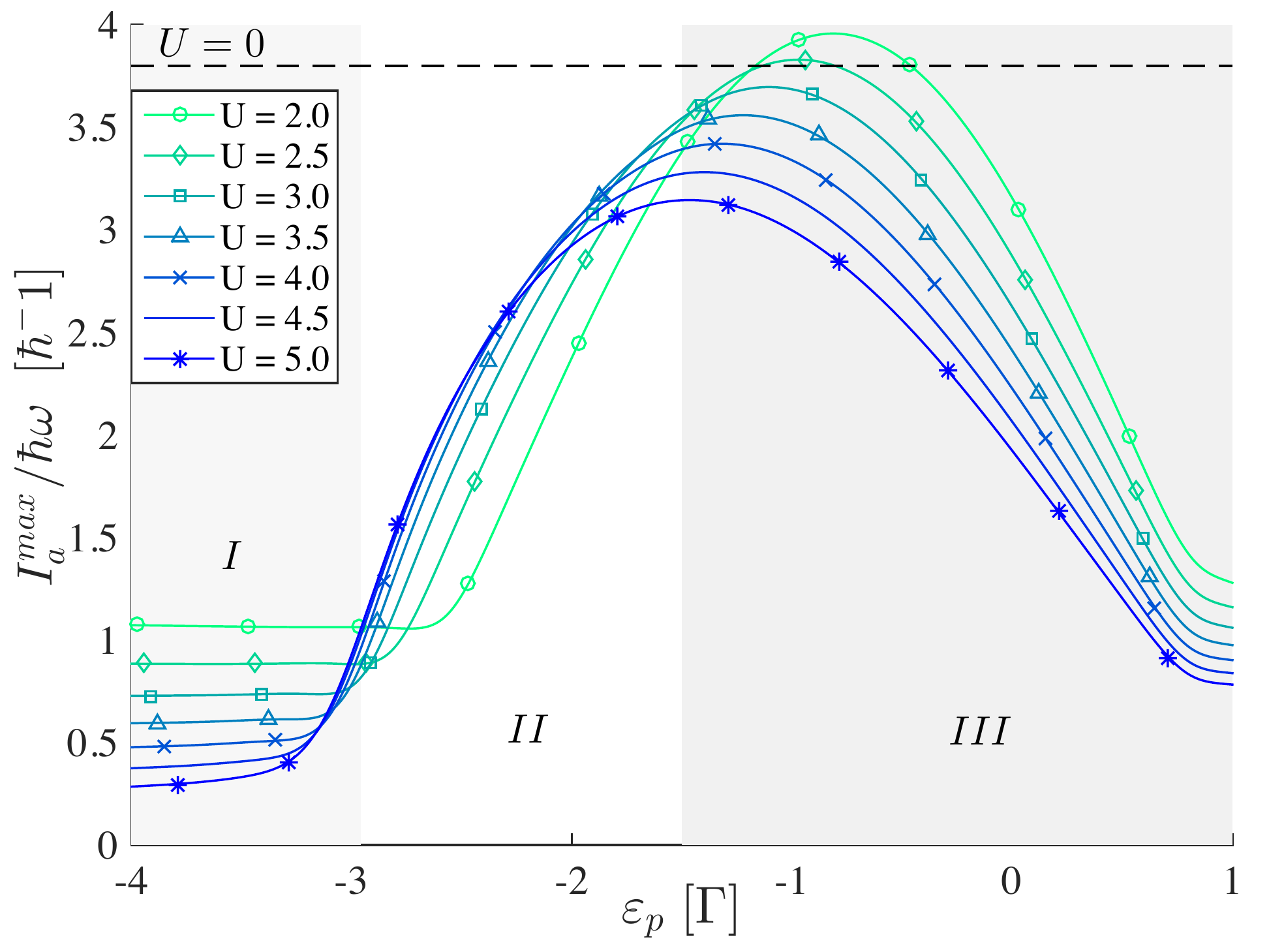}
  \caption{Maximum value of the linear response coefficient of the charge current entering the active reservoir $I^{max}_a/\hbar\omega$, as a function of $\ve_p$ and for different values of the Coulomb interaction $U$. The black dashed-line corresponds to the maximum value at $U=0$ when the quantum dots are decoupled from each other. Other parameters are the same as in Fig. \ref{fig6}.
  }\label{fig4}
\end{figure}

\begin{figure}
 \includegraphics[width=0.5\textwidth]{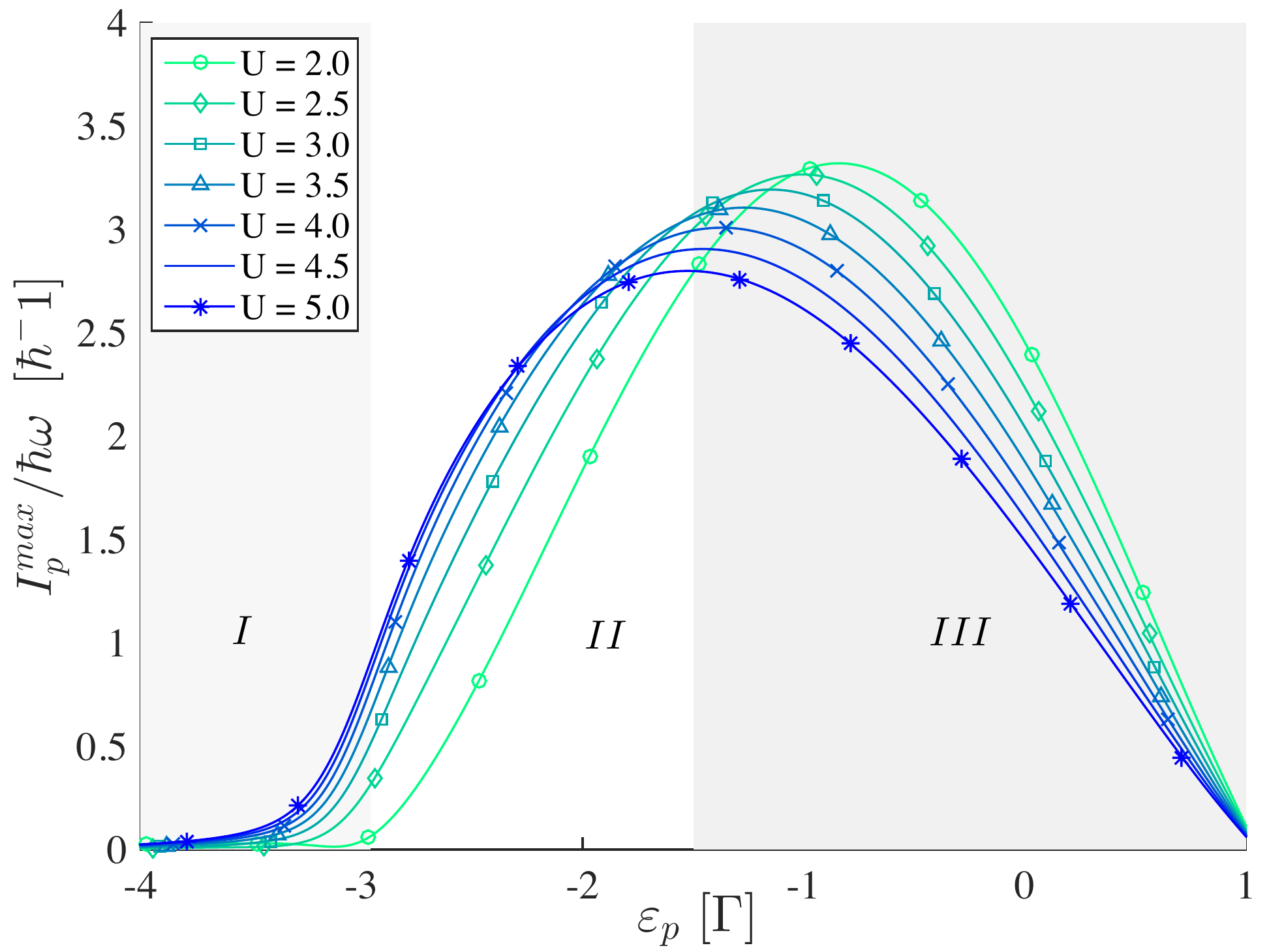}
  \caption{Maximum value of the linear response coefficient of the charge current entering the passive reservoir $I^{max}_b/\hbar\omega$, as a function of $\ve_p$ and for different values of the Coulomb interaction $U$. Other parameters are the same as in Fig. \ref{fig6}.
  }\label{fig5}
\end{figure}

In what follows we analyze the behaviour of the maximum values of the pumped charge currents $I^{max}_\alpha\equiv\mbox{max}\{\vert I^{pump}_\alpha(t)\vert\}$. We can see from Figs. \ref{fig4} and \ref{fig5} that, as mentioned above, $I_p^{max}<I_a^{max}$ for any energy $\ve_p$. Both the maximum values $I^{max}_a$ and $I^{max}_p$ exhibit a broad peak that is centered around the critical energy $\ve_p^c(U)$,
whose width extends over an energy range where the passive dot is able to exchange electrons with the right reservoir since $\vert{\ve_p^t}-\mu\vert\lesssim {\Gamma_p^t}$ for some time-intervals in the oscillation period. On the contrary, when the un-driven dot is deep below the Fermi energy $\ve_p\ll\mu$, and in the limit $\ve_p\rightarrow \Gamma$, the current in the passive dot is suppressed since $\vert{\ve_p^t}-\mu\vert> {\Gamma_p^t}$ for all time, while $I_a^{max}$ is still finite but much smaller than the maximum value at $U=0$ (dashed line in Fig. \ref{fig4}). As explained before, the reduction of the current $I_a^{max}$ in the active reservoir is due to the mutual friction between inter-dot electrons. Thus, in order to increase the current entering the active reservoir, the passive dot should be able to empty. This latter explains the fact that both currents exhibit the broad peak within the same range of energies. 
%%%%%%%%%%
 %%%%%%%%%%
 
 %%%%%%%%%%
 %%%%%%%%%%

Going more in detail, we identify three different regimes (or regions) in the behaviour of the maximum currents:
\subsubsection{Region I: for ${\ve_p}\leq-\ve_0$}

This is the light gray region in Figs. \ref{fig4} and \ref{fig5}, which is characterized by having off-peak currents. Then, $I_p^{max}\rightarrow 0$ is practically suppressed and $I_a^{max}$ is quite reduced with respect to its non-interacting value. Here, lower values of $U$ favor the generation of a charge current entering the active reservoir and make $I_p^{max}$ vanish, that is reasonable since charge variations in the passive dot are merely induced by the Coulomb coupling. The upper limit of the region was defined as the energy where there is a change in the behaviour of $I_a^{max}$ in a way that curves for a more intense interaction start to be above those corresponding to lower values of $U$, precisely at the intersection $I^{max}_a(U=2\Gamma)=I^{max}_a(U=5\Gamma)$.  

Within this range of energies, the level of the passive dot is deep below the Fermi energy and therefore the dot is filled  $\overline{\langle n_p^*\rangle}\rightarrow 1$.
Then, to allow the electron in the passive dot to tunnel into the reservoir on the right, it has to receive some energy from the active dot, that is just a portion of the power developed by the time-dependent source $\ve_a(t)$ (see Eq. (\ref{energydistribution})). 
However, the fact that $\vert\ve_p\vert>\ve_0$ renders the external source $\ve_a(t)$ unable to develop enough power to empty the passive dot, and that is why the current $I_p^{max}$ vanishes and $I_a^{max}$ flattens. As $\ve_p\rightarrow -\ve_0$ and the Coulomb coupling gets more intense by rising up $U$, 
an exchange of energy between the two dots becomes possible and $I_p^{max}$ starts to increase and so does $I_a^{max}$.

In this region, the ordering of the curves as $U$ is varied can be explained in terms of the mean energy shift $\overline{\lambda^t}$, that was previously analyzed. As the value of $U$ rises up, the Lagrange multiplier $\overline{\lambda^t}$ increases. Therefore, the effective energy of the passive dot gets closer to $\mu$ (see Fig. \ref{fig6}), which makes $I_p^{max}$ increase. On the contrary, the effective active level is shifted farther away from $\mu$ as $U$ gets larger (from Fig. \ref{fig6}, notice that  $\overline{\ve_a^t}=\overline{\ve_p^t}-\ve_p=\overline{\lambda^t}>0$), so that the level is out of resonance which consequently reduces the current entering the active reservoir.

\subsubsection{Region II: for $-\ve_0<{\ve_p}\leq \ve_p^c$}

This is the intermediate region in the Figs. \ref{fig4} and \ref{fig5}. Its upper limit is approximate and it was defined as the average critical energy $\langle\ve_p^c\rangle$ over all the values of $U$ we take, which coincides with the intersection $I_p^{max}(U=2\Gamma)=I_p^{max}(U=5\Gamma)$. 

Within this region the effective passive dot lies on average below the Fermi level of the reservoirs $\overline{\ve_p^t}\leq 0$, but it is in resonance for some time 43intervals during the oscillation period $\vert \ve_p^t-\mu\vert\lesssim\Gamma_p^t$ so that electrons can be exchanged between the passive dot and its reservoir. 
The time-dependent Lagrange multiplier $\lambda^t$ makes the passive energy level oscillate around a mean value that gets closer to $\mu$ as the interaction $U$ increases. This settlement enhances the current entering the passive reservoir, and therefore in Fig. \ref{fig5} we can see a growth of $I_p^{max}$ as $U$ rises up.
Regarding the current in the active reservoir, Fig. \ref{fig4} shows that also $I_a^{max}$ increases with $U$. Here,
the Lagrange multiplier moves up the active level as well, but in this case its mean value moves away from the Fermi level as the repulsion $U$ increases. In this sense, we would expect weaker interactions to enhance the current $I_a^{max}$ since the mean effective active level $\overline{\ve_a^t}=\overline{\lambda^t}$ would be closer to the resonance $\mu$. Nevertheless, there is another important factor that prevails and determines the behaviour of the $I_a^{max}$ within this region, and it is the following. For an electron to be pumped into the active dot, the passive dot should empty, since the Coulomb repulsion prevents (less or more, depending on the value of $U$) from the double occupancy of the double-dot system. For achieving that, the connection between the two dots should be reinforced by increasing the Coulomb coupling $U$ so that they can exchange more energy which induce the pumping of charge in the passive dot.   

\subsubsection{Region III: for ${\ve_p}> \ve_p^c$}
The last region is distinguish by the fact that both the effective levels oscillates in time around a mean value that is above the Fermi level $\mu=0$, so $\overline{\ve_a^t}\geq 0$ and $\overline{\ve_p^t}\geq 0$ (see Fig. \ref{fig6}). 
Therefore, for the level of the active dot as well as for the passive one, larger values of $U$ imply being further above the Fermi level of the reservoirs so that when the levels move away from the resonance both currents $I_a^{max}$ and $I_p^{max}$ decrease. Hence, in Figs. \ref{fig4} and \ref{fig5}, curves for lower values of $U$ are above those corresponding to larger interactions. 
As the mean effective passive level goes further above the chemical energy, the current $I_p^{max}$ decreases and so does $I_a^{max}$, since charge transport in the active subsystem strongly depends on the possibility of pumping electrons in the passive subsystem.
In the limit $\ve_p\rightarrow\Gamma$ the effective passive level is off-resonance at all times, since $\vert\ve_p^t-\mu\vert>\Gamma_p^t$ and consequently the current $I_p^{max}\rightarrow 0$ almost vanishes. As mention before, the mean occupancy of the passive dot in this limit is not completely zero so that there is still a mutual friction between inter-dot electrons that reduces the maximum current entering the active reservoir with respect to the $U=0$ case. 
Naturally, the Coulomb coupling between the dots has a less impact on the active current as the passive level overcomes $\mu$ and it empties. That is why  the active current $I_a^{max}$ seems to stabilize in this limit at higher values than when $\ve_p\ll\mu$.

On the other hand, we can see that when $U<\ve_0$ the peak in the active current $I_a^{max}$ exceeds the non-interacting value. This aspect has to do with the power supply from the external driving source, and it will be discussed in the following section.   
   
\subsection{Energy fluxes}

In this section we study the energy that is dissipated deep inside the reservoirs of the subsystems, $\dot{Q}_a$ and $\dot{E}^{a\rightarrow p}$, whose expressions were presented in Eq. (\ref{powerdiss}). When evaluating Eqs. (\ref{curr}) and (\ref{powerdiss}) at zero temperature, an instantaneous Joule Law relation emerges 
\be\label{joule}
P_\alpha^{diss}(t)=R_q (I_\alpha^{pump}(t))^2 \;\;\;\text{at $T=0$,}
\ee
with an universal resistance $R_q=h/2{e^2}$ that is the charge relaxation resistance found in Refs. \cite{singledot,res1} and also observed in Ref. \cite{res2}.
In particular, for the passive subsystem $\alpha=p$, the above equation reads $\dot{E}^{a\rightarrow p}=R_q(I_p^{pump})^2$ and shows how the 
transfer of energy between the subsystems (from the active subsystem to passive one) is accompanied by the induction of charge pumping in the un-driven dot. 

\begin{figure}[h]
 \includegraphics[width=0.5\textwidth]{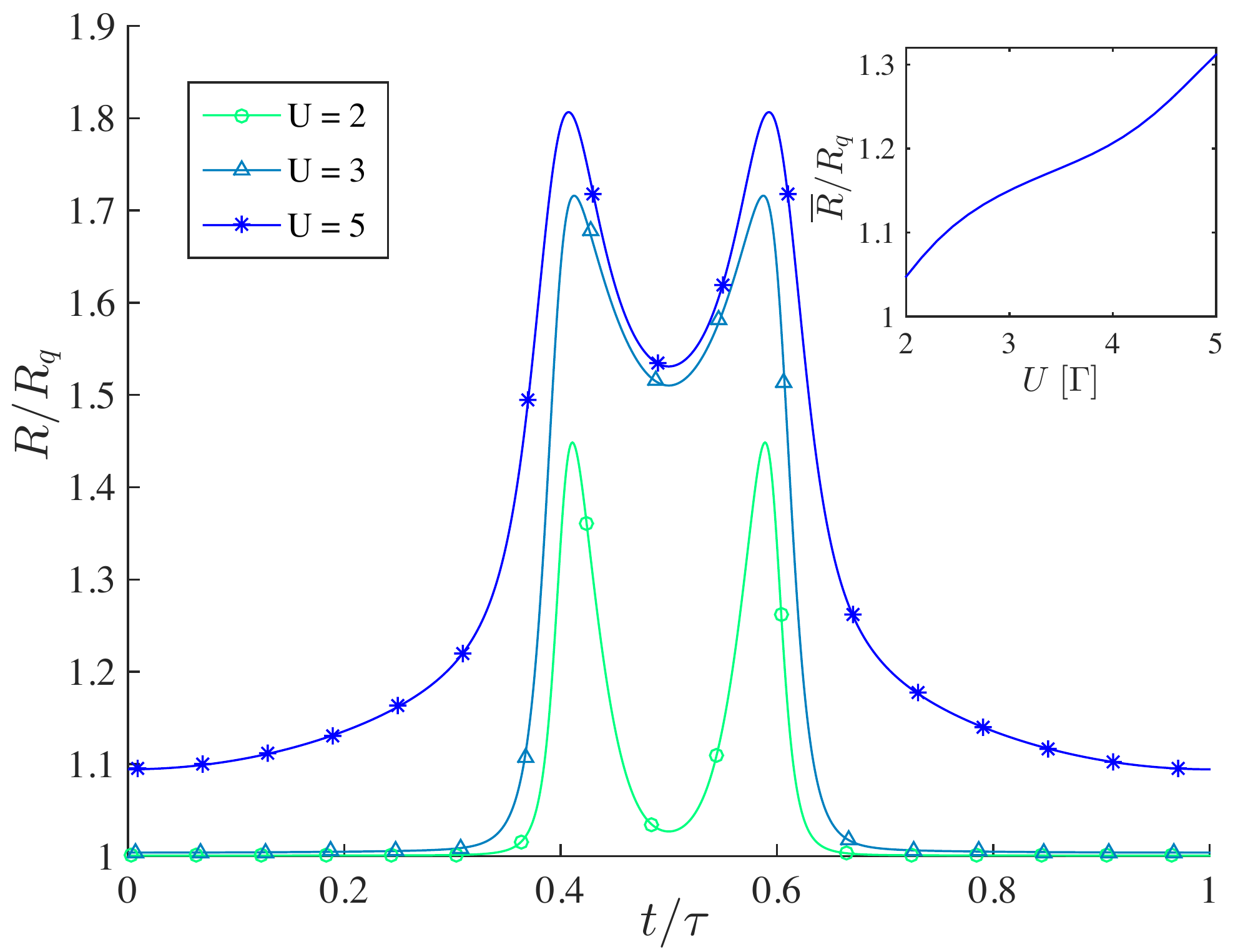}
  \caption{Effective resistance $R(t)$ in units of $R_q=h/2{e^2}$  as a function of time. Parameters are: $\ve_p=-2.5\Gamma$, $\mu=0$, $T=0$, $\ve_0=3\Gamma$. Inset: mean value of the resistance $\overline{R}$ over one driving period $\tau$ as a function of the interaction $U$. All the energies are in units of $\Gamma$.
  }\label{fig8}
\end{figure}

Similarly, when $\alpha=a$, Eq. (\ref{joule}) relates the heat $\dot{Q}_a$ that is dissipated in the active reservoir to pumping in the driven dot. However, unlike in the passive dot, in the active subsystem not all the energy that the active dot receives from the external ac source is then dissipated as heat in the active reservoir. Here, as shown in Eq. (\ref{energydistribution}), dissipation corresponds just to a portion of the power delivered by the external source $P^{ac}_{diss}$, while the rest $\dot{E}^{a\rightarrow p}$ is transferred to the passive subsystem in order to ``palliate" the Coulomb friction. This does not happen when the two quantum dots are uncoupled, i.e. for $U=0$, since in that case the dissipated heat $\dot{Q}_a\vert_{U=0}=R_q (I_a^{pump})^2\vert_{U=0}=P^{ac}_{diss}\vert_{U=0}$ is equal to the ac power because there is no flow of energy to the passive dot. Thus, to generate a certain charge current in the active subsystem at finite $U$, the source has to inject a higher amount of energy in order to get over the friction.   
This fact should be reflected in the relation between $P_{diss}^{ac}$ and $I_a^{pump}$, through an effective resistance for the active dot which should be larger than the non-interacting value $R_q$. For analyzing this, we insert Eq. (\ref{joule}) in the total power, so
\be\label{joule2}
P_{diss}^{ac}(t)=P_a^{diss}(t)+P_p^{diss}(t)=R(t)[I_a^{pump}(t)]^2,
\ee
where we define
\be
R(t)\equiv R_q\left(1+\left[\frac{I_p^{pump}(t)}{I_a^{pump}(t)}\right]^2\right)
\ee
as the effective resistance, that is a manifestly positive quantity at all times. Therefore, we can see Eq. (\ref{joule2}) as a Joule law with an instantaneous effective resistance $R(t)$ for the total energy dissipation due to pumping in the active dot. Here, we can notice from Eqs. (\ref{joule}) and (\ref{joule2}), that the overcome of the non-interacting maximum current in Fig. \ref{fig4} for $\ve_0>U$ is just because the source is injecting more energy with respect to the $U=0$ case.
As an example, we show in Fig. \ref{fig8} the behaviour of the effective resistance $R$ for different values of the interaction $U$ when the passive level is filled $\ve_p=-2.5\Gamma$. We can see that the effective resistance fulfills the relation $R(t)\geq R_q$ at all times, and that it is not universal since it depends on the interaction $U$ and the spectral properties of the dots. As expected, the effective resistance in the active dot due to the presence of an electron in the passive dot gets larger as $U$ increases.
%%%%%%%%%%
 %%%%%%%%%%
 
Finally, we study the efficiency of the energy transfer between the two dots over a cycle of the external driving source. This quantity measures the amount of heat leaking from the active to the passive system. The knowledge of its behavior can be crucial to guide the  design of experimental set-ups. A first natural use is to monitor this quantity in order to prevent an excessive heating of the passive part or, on the other hand, in order to use the leaking heat as an energy input (work), for example in positions where an external source can not be easily connected.

\begin{figure}[h]
 \includegraphics[width=0.5\textwidth]{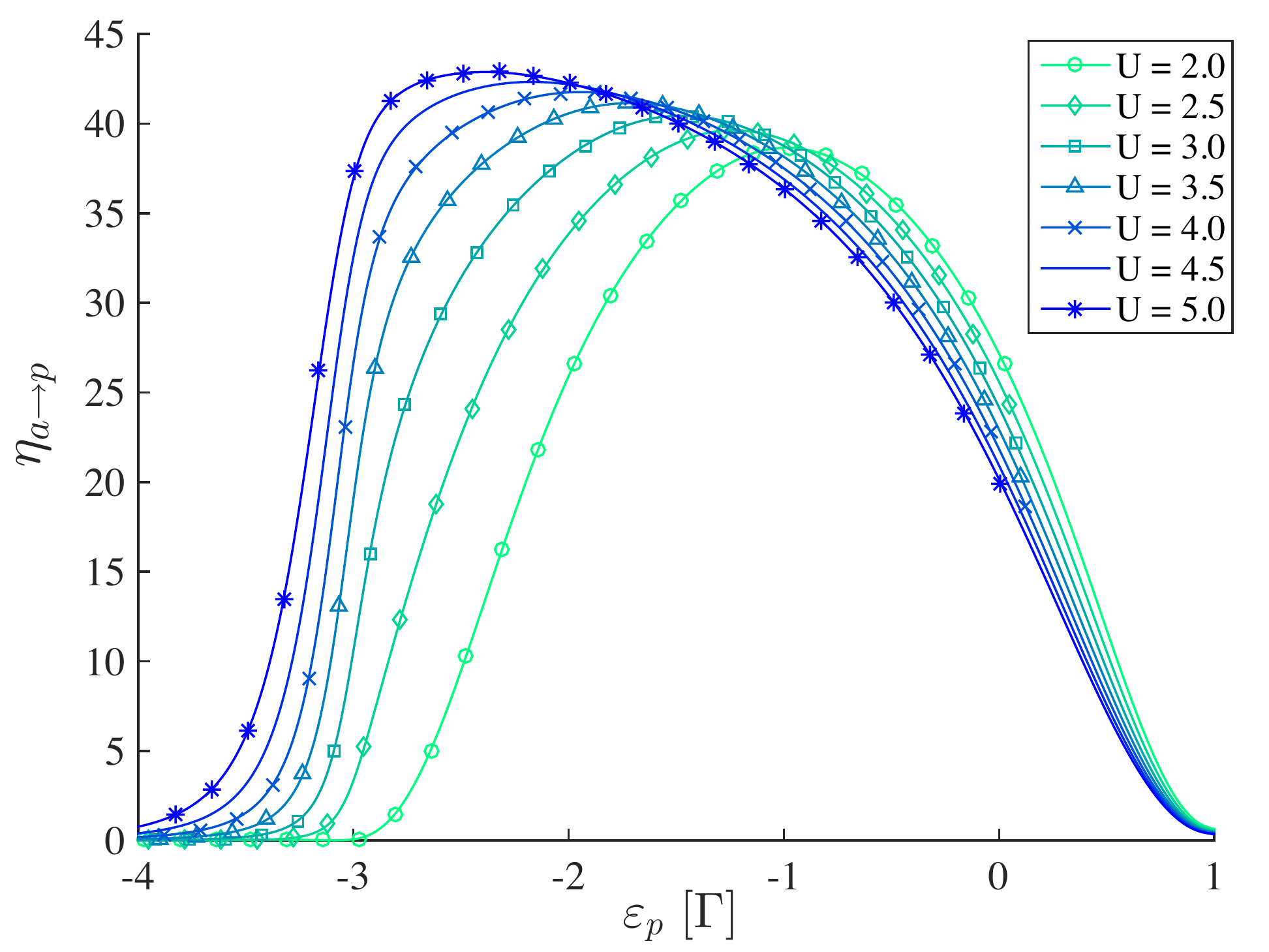}
  \caption{Efficiency of the energy transfer from the active quantum dot to the passive one $\eta_{a\rightarrow p}=100\times{\overline{\dot{E}^{a\rightarrow p}}}/\overline{P^{ac}} $, as a function of $\ve_p$. Parameters are the same as in Fig. \ref{fig6}. 
  }\label{fig7}
\end{figure}

For that, we define the efficiency as 
\be
\eta_{a\rightarrow p}=\frac{\overline{\dot{E}^{a\rightarrow p}}}{\overline{P^{ac}}}\times 100,
\ee
that is the percentage of the averaged injected power $\overline{P^{ac}}$ that is transmitted to the passive quantum dot. Fig. \ref{fig7} shows the results as function of the level of the passive dot $\ve_p$, again for $\mu=0$ and $T=0$. Naturally, the broad peak in $\eta_{a\rightarrow p}$ occurs within the same range of energies as the peak of the charge pumping in the passive reservoir. We can see that the efficiency also follows the behaviour of $I_p^{pump}$ as $U$ is varied. In the sense that larger interactions improve the energy transmission when the effective passive level is filled on average $\ve_p<\ve_p^c$, while they make it worse when the level is above the Fermi sea $\ve_p>\ve_p^c$.

Surprisingly we find that a maximum of around $\sim 43\%$ (maximum of $\eta_{a\rightarrow p}$ when $U=5\Gamma$) of the energy delivered by the ac source is transmitted to the passive dot, which is quite high. As it was mentioned before, this amount of transmitted energy is then delivered to the passive reservoir and dissipated there as heat. However, this energy could be eventually transformed into useful work in a proper designed setup.   

 %%%%%%%%%%
 %%%%%%%%%%
 
\section{Conclusions}\label{conclusions}

In this work we have studied the transport of charge and energy in two Coulomb-coupled quantum dots. We considered the setup in Fig. \ref{fig1} in which only one of the two dots (the active dot) is adiabatically driven by a time-periodic gate, while the other quantum dot remains undriven (passive dot). We have shown that although the Coulomb coupling does not allow for electron transfer between the two quantum dots, it enables a transfer of energy between them (sketched in Fig. \ref{fig2}) that eventually induces a pumping of charge also in the undriven dot, even at zero temperature. In order to treat the  effects of the Coulomb interaction at low temperatures, we used the time-dependent slave-spin 1 formulation within mean-field presented in Ref. \cite{slaveapin}, which turns out to be advantageous for describing the induction of transport in the passive subsystem due to the Coulomb repulsion, since it is simply explained in terms of effective time-dependent driving  fields acting on the passive quantum dot.

We have found that the pumping currents that are induced in the passive dot due to the mutual friction are of the same order, even if always smaller, than those generated in the driven dot. Moreover, we identify three different regimes in the behavior of the charge fluxes as a function of the energy level of the passive dot.

As far as energy transport is concerned, we have found that the dissipation in both the reservoirs due to pumping at zero temperature is given by a Joule law controlled by the universal charge relaxation resistance $R_q$. In addition, we also derived a Joule law for the total energy dissipation due to pumping in the active dot, which is now controlled by an instantaneous non-universal resistance $R(t)$. We have derived that $R(t)\geq R_q$ at all times and increases with the interaction $U$.
Finally, we analyzed the efficiency of the energy transfer from the active dot to the passive one, and our results showed that for suitable parameters we can reach a regime where up to $\sim 43\%$ of the energy delivered by the ac source to the active dot is transmitted to the passive dot. 

Our results represent a significant advance toward a full understanding of drag effects of charge and energy in Coulomb coupled quantum dots systems which is expected to have potential implications for nanoelectronics. 
The development of low-dimensional electronic devices least to set-ups where the components are located very close to each other. This leads inevitably to an increased Coulomb interaction, which can have undesired consequences. For example currents can be induced in undesired regions of the device which do not contribute to its functionality, and heat can be accumulated or dissipated in inappropriate locations thus ruining or even disrupting the material.

Our results provide a novel piece of information about the conditions favoring such undesired phenomena which can help to avoid or reduce their impact. While the current calculations refer to a very simplified set-up, they open the path to calculations for more realistic descriptions of the devices owing to the simplicity and the reliability of our theoretical approach.

\section{ACKNOWLEDGEMENTS}
We acknowledge support from the Italian Ministry of University and Research (MIUR) through the PRIN2017 project CEnTral (Protocol Number 20172H2SC4). We also thank Fabio Caleffi for useful discussions.

\end{document}